\documentclass[conference]{IEEEtran}
\pdfoutput=1
\IEEEoverridecommandlockouts

\usepackage{tikz-network}

\usepackage{amsmath}
\usepackage{graphicx}

\usepackage[colorlinks=true,allcolors=blue]{hyperref}
\usepackage{amsthm}

\usepackage{algorithm}
\usepackage{pseudo}
\usepackage[most]{tcolorbox}

\usepackage{lipsum}

\theoremstyle{definition}
\newtheorem{definition}{Definition}[section]

\theoremstyle{definition}

\newtheorem{usecase}{Use case}[section]

\usepackage{xspace}
\newcommand{\SSC}{SSC\xspace}

\title{Assessing the Threat Level of Software Supply Chains with the Log Model}
\author{
  \IEEEauthorblockN{Luís Soeiro\IEEEauthorrefmark{1}, Thomas Robert\IEEEauthorrefmark{1}, Stefano Zacchiroli\IEEEauthorrefmark{1}}
  \IEEEauthorblockA{\IEEEauthorrefmark{1}LTCI, Télécom Paris, Institut Polytechnique
		 de Paris, France\\
    \{luis.soeiro, thomas.robert, stefano.zacchiroli\}@telecom-paris.fr}
  \thanks{Supported by the industrial chair Cybersecurity for Critical Networked Infrastructures (cyberCNI.fr) with support of the FEDER development fund of the Brittany region, France.}
}

\begin{document}
\maketitle

\begin{abstract}
The use of free and open source software (FOSS) components in all software
	systems is estimated to be above 90\%. With such high usage and because of the
	heterogeneity of FOSS tools, repositories, developers and ecosystem, the level
	of complexity of managing software development has also increased. This has
	amplified both the attack surface for malicious actors and the difficulty of
	making sure that the software products are free from threats. The rise of
	security incidents involving high profile attacks is evidence that there is
	still much to be done to safeguard software products and the FOSS supply chain.

	Software Composition Analysis (SCA) tools and the study of attack trees help
	with improving security. However, they still lack the ability to
	comprehensively address how interactions within the software supply chain may
	impact security.

	This work presents a novel approach of assessing threat levels in FOSS supply
	chains with the log model. This model provides information capture and threat
	propagation analysis that not only account for security risks that may be
	caused by attacks and the usage of vulnerable software, but also how they
	interact with the other elements to affect the threat level for any element in
	the model.

\end{abstract}


\section{Introduction}
The security of technological supply chains is a hot topic especially for
software, with a growing amount of high-profile attacks~\cite{Ohm2020,
	Ladisa2022}, like the highly-publicized one on SolarWinds product
Orion~\cite{Peisert2021}. While security of supply chains for manufacturers
have been deeply studied for years, the way software is produced, stored, and
distributed adds different challenges to the security of supply chains for
software~\cite{ghadge2020managing,hammi2023security}. Most software systems
depend on many third party free and open source software
artifacts~\cite{Harutyunyan2020,Vu2021} that have been copied, transformed, or
combined to form the executable software product. Each of those artifacts, in
turn, might be a combination of other software artifacts, and so on,
recursively forming the software supply chain for the final software product.
If any of those components is vulnerable or compromised there is a possibility
that the end software products of the related \emph{Software Supply Chain
	(\SSC)} may be be affected.

The study of software dependencies (e.g., source code, libraries, other
components) has been the main focus of software development
practitioners~\cite{Abate2020DepSolving} and empirical
researchers~\cite{Decan2019DepNetworks}. Such analyses target different goals:
to check for license compliance issues~\cite{German2017LicenseInconsistencies},
or to determine the vulnerability of a software product. Software Composition
Analysis (SCA) tools~\cite{Ombredanne2020} usually help by generating a list of
dependencies for a package and then searching for vulnerable or compromised
components that were included with it \cite{Imtiaz2021}. The purpose of these
analyses is to propagate knowledge about dependencies up to the end product.
Yet, propagating security threat information only from components identified as
``parts'' of the end product is not satisfactory. In \cite{Ladisa2022}, authors
pointed out that there was mainly two strategies to compromise a software
product: compromising the host of the build process, or compromising the
storage and distribution infrastructure of software artifacts.

The presence of vulnerabilities in the operational part of the software supply
chain can make it possible for an attacker to compromise software artifacts in
the build environment used during builds. Compromised tools executed to carry
out required steps such as compilation, linking or packaging of the software
are even more concerning in terms of security than a vulnerable library as they
may contain code that runs at build time to inject delayed attacks. The already
mentioned SolarWinds'Orion attack~\cite{Peisert2021} is a perfect example of
this.

In terms of threat modeling and attack scenario description, the state of the
art of research is to rely on attack trees~\cite{Schneier1999AttackTrees}. This
formalism is generally used to describe attacks or threat for classes of
systems without accounting for any specific knowledge of it. Hence, there is a
lack of formalism that integrates insights about vulnerable or compromised
software artifacts, and compromised hosts that run build processes or
distribute packages to determine the threat level represented by a piece of
software.

\paragraph*{Paper contributions}

We propose the \textit{log model} as the basis for an approach to keep a
history of software supply chain activities for a given software product, which
allows practitioners to gather a better picture of the threat level represented
by a software component. We then propose an algorithm to calculate the
resulting propagated security status for any element of the supply chain
captured by the model. Such approach is complementary to direct detection
procedure of malicious content in software product, but may spare a lot a time
identifying the components that are most likely to actually contain malicious
code or the most critical vulnerabilities.

\paragraph*{Paper structure}

Section~\ref{sec:rw} details related works and motivating literature relevant
for our proposal. Section~\ref{sec:approach} details the approach and the core
elements needed to be able to carry the above mentioned analysis.
Section~\ref{sec:thtmodel} details the threat levels for each entity and gives
the intuition about how it can propagate down to the end products of the \SSC.
Section~\ref{sec:logmodel} presents the model used to capture how the software
supply chain has been used to produce a given end product, or a set of related
products. From those elements, section~\ref{sec:analysis} details the proposal
for a threat level propagation analysis that can be used to grasp first
insights on \SSC vulnerabilities. It can also be used to determine which
elements of the \SSC might be compromised when additional knowledge about
compromised artifacts, execution or storage resources is provided. Such cases
are illustrated in section~\ref{sec:usecases}. Finally,
section~\ref{sec:conclusion} presents the conclusions.

\section{Related Work}
\label{sec:rw}

Core principles about supply chains security have been studied and led to the
identification of abstract threat types~\cite{hammi2023security} that apply to
supply chains in general and to \SSC specifically. Yet, we are focusing on
detailed threat modeling, which is most often part of Cyber-Threat Intelligence
and is close to detection concerns~\cite{8240774}. Attack
trees~\cite{Schneier1999AttackTrees} and attacks graphs are two complementary
tools for identifying and describing attack requirements and to reason about
attack propagation.

The most up to date attack tree in this field is provided by Ladisa et
al.~\cite{Ladisa2022} extending previous work by Ohm et al.~\cite{Ohm2020}. It
identifies three main vectors to carry out attacks by compromising or abusing:
the way software source code is contributed and injected into software supply
chains, build environments, or the distribution infrastructure of software
(e.g., package managers, repositories). Hence, we cover two out of three of the
main attack vectors, but the coverage of repositories is less clear as the
attack scenarios most often do not assume compromised repositories but abuses
of the mechanisms and bad practices when it comes to consume software. This
point of view is consistent with the general understanding of the core issues a
SSC has to cope with \cite{7180277}.

Examples of attack graphs integrating attacks steps on the supply chain can be
found in \cite{duman2019modeling}. The advantage of attack graphs over attack
trees is that they should account for the actual system for which one wants to
determine whether a piece of software is a threat. The main issue of this
contribution is that the knowledge of compromised software artifacts should be
an input of the graph model. Instead, the graph simply accounts for such
events, not helping to see how such events propagate into software supply
chains.

Other graph-based strategies have been proposed to better identify or detect
vulnerabilities by trying to understand changes in versions and evaluating how
they change the semantics to detect
vulnerabilities~\cite{yamaguchi2014modeling, 10123571}. Such approaches are
complementary to ours. They require actual access to the source code used in
the \SSC and do not provide a clear strategy to study the impact of a software
component vulnerability on the \SSC, they only account for vulnerability of
source packages. Their advantage is then to avoid too pessimistic conclusions
by considering which ones are actually reachable during end product potential
execution (e.g., thanks to call graph analysis). A vulnerable software in our
case can have two impacts: i) introduce directly the same vulnerability in the
product; ii) make some step of the \SSC vulnerable and open the door to the
production of a compromised end product. Hence, we provide less accurate
information in terms of potential compromises, but we require less information.

From the detection point of view, the dependencies of an end product on sources
are used to infer from the source code an estimate of the level of compromises
or vulnerabilities for each commit in a version control
system~\cite{perl2015vccfinder}. Despite the advantages of being able to
determine those vulnerabilities for each commit, the propagation of their
impact needs to be understood~\cite{8009930}. The latter publication shows how
detection is interesting but limited without the means to propagate the
knowledge to dependent software artifacts, build processes, execution
resources, provider-consumer links, or versions histories.

Another approach to reason about software security risks is to use Software
Bills of Materials (SBOMs) to look for vulnerable components~\cite{Zahan2023}.
The leading industry standards, SWID, SPDX, and ClycloneDX, can represent
software components and be automatically generated and consumed. However, there
are no agreements on the data fields that should be present. Also, SBOM
adoption is still lacking~\cite{Xia2023}, despite many years of availability of
the SPDX specification~\cite{stewart2010software}. Additionally, even a
comprehensive SBOM only captures a static snapshot of the dependencies for a
software product. It doesn't include information on the interactions that have
occurred among all the elements of the software supply chain during the build
processes. Those interactions may affect threat propagations and so are
necessary for our current work.

To our knowledge the existing software supply chain models and information
capture approaches suffer from a lack of requirements and semantics for the
dependencies that they can record. It does not make them incorrect but
ultimately it does make them impossible to exploit for the kind of automated
propagation analysis we target.

\section{Approach}
\label{sec:approach}
Our approach can be seen as an extension of usual recursive dependency analyses
for vulnerable software~\cite{7774522}. As pointed out in the previous section,
the outcome of a \SSC does not only represent a threat because it may
distribute vulnerable software. Since a \SSC can also be a vector to distribute
malicious software, our proposal offer a method to reason about both cases.

We introduce a dependency graph model that accounts for resources that were
used along the software supply chain (e.g., host machines, software tools) of a
given product. This model, called the \textsl{log model}, identifies different
elements involved along the supply chain that have a direct role in either
propagating vulnerabilities or injecting and distributing malicious code. From
such a model, we define a set of generic inference rules that help to identify
all the consequences of a current state of knowledge in terms of threats. These
rules allows the propagation of knowledge of vulnerable or malicious software,
and of vulnerable or malicious host machines along the dependencies among the
\SSC operations. The advantage of this approach is to combine different sources
of information by using with this notion of state of knowledge, and then to
infer the consequences of this state of knowledge for the end product.

Hence the first step is to decompose the main element types of the \SSC. At
first two types of elements are distinguished: host machines, called
\textsl{hosts}, and software elements, called \textsl{software artifacts}. It
helps us to distinguish components that execute something from those that could
be executed. We need to be able to describe the steps of the \SSC that produce
something: \textsl{transformers}. Such components allow us to capture three
distinct behaviors for software dependencies: software artifacts that carry out
the operations related to transformers (e.g., compilers, packagers), artifacts
that are integrated directly or indirectly to the product (e.g., libraries,
source code), and artifacts that are necessary for operations of hosts (e.g.,
operating systems, hypervisors, containers). Because a host can be involved in
different steps of the \SSC, it may provide for each of these transformers a
different environment in which the operation is carried out. This is the last
type we introduce, the \textsl{build environment}, to capture the exact context
in which the operations took place.

In the next section, we provide the details of the information we propose to
attach to each element, and the inference rules that can be used to capture the
interplay of software artifacts, hosts, build environments and transformers in
terms of propagation of vulnerabilities and malicious content.

\section{A Threat Model for Software Supply Chains}
\label{sec:thtmodel}

The focus of the proposed threat model is on the impact of attacks rooted
within the resources found on software supply chains. We aim to calculate the
security status of any element of the software supply chain.

\begin{definition}[Security Status Calculus]
	The resulting security status for an element in a software
	supply chain is the security status of maximum impact of all elements that
	contribute directly to it.
\end{definition}

The elements necessary to perform such analyses are presented, along its
possible security implication.

\begin{definition}[Host]\label{def:host}
	A host is a computer system which stores and uses software resources.
\end{definition}

\begin{definition}[Software Artifact]\label{def:software_artifact}
	A software artifact is a named set of software parts that are made
	available to be executed on host machines.
\end{definition}

\label{def:sw_security_state}
Let $Q_S = [ \textit{safe}, \textit{vulnerable}, \textit{malicious} ]$ be an
ordered list of possible software security states for a
\textit{software artifact}. They are arranged from the lowest to the
highest impact they may have on the system, where:

\begin{itemize}
	\item \textbf{safe} - there are no known vulnerabilities reported, so it has
	      no adverse impact on the security of the system, a priori;
	\item \textbf{vulnerable} - there is at least one known vulnerability
	      reported that makes it exploitable, but it requires at least one action
	      from the attacker to carry the exploitation;
	\item \textbf{malicious} - the software contains code that is already
	      intended to perform malicious activities in a potentially automated way
	      (e.g., open backdoors for remote
	      shell access, injection of malicious code into other artifacts).
\end{itemize}

We consider the final security status of a software artifact the result of
combining its known initial security status with the propagated security level
from other elements. For the initial state we need existing security
information related to vulnerable and malicious software artifacts. For
vulnerable software, there has been ongoing identification efforts by the
software community and to make the information searchable~\cite{Chang2011}. Let
$\chi_{SV}$ be the set of software artifacts with known vulnerabilities,
already extracted from such information sources.

Since the first self replicating software was crafted and
observed~\cite{Orman2003}, malicious software (e.g., virus, trojans) has been
studied. Such software often use evasion techniques to make them harder to be
detected~\cite{Maniriho2022}. Since we can't rely on usual software artifact
identifiers (e.g., file name, version, author) or even on intrinsic ones (e.g,.
cryptographic hashes), malicious software databases are not feasible. One must
instead rely on file system scanning and intrusion detection techniques (e.g.,
cryptographic hash alteration). Let $\chi_{SM}$ be the set of malicious
software artifacts detected or inferred on the system by any means.

In the current work, we assume a worst case scenario: when a host is known to
have being compromised, all software artifacts that come from it are considered
to be malicious. However, there is an exception. There might be an
uncompromised host (e.g., no attacks were reported or observed) where a
malicious software was published to be distributed, but was not itself
executed. If this malicious software artifact is in the supply chain for other
software artifacts, this host will be considered compromised for that supply
chain, because it is distributing compromised software artifacts. If, on the
other hand, another safe software artifact is fetched from this same
uncompromised host, as part of another software supply chain, it won't be
classified as malicious for that software supply chain.

\label{def:hw_security_state}
Let $Q_H = [ \textit{safe}, \textit{vulnerable}, \textit{compromised} ]$ be an
ordered list of possible software security states for a \textit{host}.
The list is arranged from the lowest to the highest impact they may have on the system, where:
\begin{itemize}
	\item \textbf{safe} - all software artifacts present on the system
	      have the \textit{safe} security status and there are no reports
	      of security breaches of the host system;
	\item \textbf{vulnerable} - there is at least one software artifact present
	      in the host that has the \textit{vulnerable} security status, but none
	      that has the \textit{malicious} security status,
	      and there are no reports of security breaches of the host system;
	\item \textbf{compromised} - there is at least one software artifact
	      present in the host that has
	      the \textit{malicious} security status or there is at least one
	      report of a security breach for the host.
\end{itemize}

Let $Status_S$ be the result of combining the security status of all
\textit{software artifacts} that were observed to be present in a \textit{host}
during the execution of build activities. We combine them with the initial
security status of the \textit{host} to obtain its security result.

In order to obtain the initial security status of host, it might be necessary
to rely on external attack disclosures or intrusion detection
techniques~\cite{Liao2013}. We assume such information is already available.
Let $\chi_{HV}$ be the set of hosts that are known to be vulnerable and
$\chi_{HM}$ be the set of hosts that are known to have been compromised. For a
host $h$, let $Status_H$ be \textit{malicious} if $h \in \chi_{HM}$,
\textit{vulnerable} if $h \notin \chi_{HM} \land h \in \chi_{HV}$, and
\textit{safe} otherwise. Table \ref{table:rules_host} presents the possible
combinations of resulting security status for a \textit{host}.

\begin{table}
	\caption{Rules for evaluating the current security status for a host}
	\label{table:rules_host}
	\centering
	\begin{tabular}[pos]{ |c|c|c|}

		\hline
		$\boldsymbol{Status_S}$ & $\boldsymbol{Status_H}$ & \textbf{Host result} \\
		\hline
		safe                    & safe                    & safe                 \\
		\hline
		vulnerable              & safe                    & vulnerable           \\
		\hline
		malicious               & safe                    & compromised          \\
		\hline
		safe                    & compromised             & compromised          \\
		\hline
		vulnerable              & compromised             & compromised          \\
		\hline
		malicious               & compromised             & compromised          \\
		\hline
	\end{tabular}
\end{table}

Besides providing storage and distribution functionalities for storage
artifacts, hosts also provide an environment to execute \textit{transformers}.

\begin{definition}[Transformer]\label{def:transformer}
	A transformer is a step of the software supply chain where specified
	software artifacts are executed to carry out operations (e.g.,
	fetching, compiling, packaging) on other software artifacts to
	generate one or more related software artifacts.
\end{definition}

A \textit{transformer} has two distinct types of input software artifacts:
those that carry out the execution and those that are to be acted upon. It has
just one kind of output, the software artifacts that were generated.

The security status of a \textit{transformer} is calculated in two phases. Let
$B_{tools}$ be the set of \textit{software artifacts} that are carry out
operations and $A$ the set of all other \textit{software artifacts}. In the
first phase we calculate the combined security status of $B_{tools}$ and $A$
according to the table \ref{table:rules_transformer1}, where column ``Build
tools'' shows the combined security of $B_{tools}$, column ``Other inputs''
shows the combined security of \textit{software artifacts} that are integrated
directly or indirectly to the generated artifacts, and column ``T1 result''
shows the security status for the phase 1 of transformer calculus. The second
phase will be presented later.

\begin{table}
	\caption{Rules for transformer phase 1}
	\label{table:rules_transformer1}
	\centering
	\begin{tabular}[pos]{ |c|c|c|}

		\hline
		\textbf{Build tools} & \textbf{Other inputs} & \textbf{T1 result} \\
		\hline
		safe                 & safe                  & safe               \\
		\hline
		vulnerable           & safe                  & safe               \\
		\hline
		malicious            & safe                  & malicious          \\
		\hline
		safe                 & vulnerable            & vulnerable         \\
		\hline
		vulnerable           & vulnerable            & vulnerable         \\
		\hline
		malicious            & vulnerable            & malicious          \\
		\hline
		safe                 & malicious             & malicious          \\
		\hline
		vulnerable           & malicious             & malicious          \\
		\hline
		malicious            & malicious             & malicious          \\
		\hline
	\end{tabular}
\end{table}

\begin{definition}[Build environment]\label{def:build_environment}
	An build environment is a computer system in which software artifacts can
	be executed while performing activities related to generating other software
	artifacts.
\end{definition}

One host can provide a number of build environments, possibly concurrently, for
preparation of different software artifacts in different supply chains. The
security status of a \textit{build environment} is the combination of the
security statuses of its \textit{host} and the \textit{software artifacts} that
were present during the build.

Vulnerable software artifacts that performed a supporting role for the
\textit{build environment}, i.e., operating system components, don't propagate
vulnerabilities to the \textit{transformer} generated artifacts, because they
can't directly or indirectly modify the generated software artifacts.

On the other hand, when a malicious software is present in a \textit{build
	environment}, we consider that it may have the ability to compromise other
software artifacts, including the \textit{transformer} components. For a
\textit{build environment}, we use the same set of possible security states of
\textit{hosts}, and its security status is a result of the security status of
its underlining \textit{host} combined with the security status of all input
\textit{software artifacts} that there were present. The rules are shown in the
table \ref{table:rules_build_environment}. The columns ``Host status'', ``Input
artifacts'', and ``B.env result'' show respectively, the current security
statuses for the \textit{host}, input \textit{software artifacts}, and the
resulting \textit{build environment}.

\begin{table}
	\caption{Rules for build environment security calculus}
	\label{table:rules_build_environment}
	\centering
	\begin{tabular}[pos]{ |c|c|c|}

		\hline
		\textbf{Host status} & \textbf{Input artifacts} & \textbf{B.env result} \\
		\hline
		safe                 & safe                     & safe                  \\
		\hline
		vulnerable           & safe                     & safe                  \\
		\hline
		malicious            & safe                     & compromised           \\
		\hline
		safe                 & vulnerable               & safe                  \\
		\hline
		vulnerable           & vulnerable               & safe                  \\
		\hline
		malicious            & vulnerable               & compromised           \\
		\hline
		safe                 & malicious                & compromised           \\
		\hline
		vulnerable           & malicious                & compromised           \\
		\hline
		malicious            & malicious                & compromised           \\
		\hline
	\end{tabular}
\end{table}

Finally we compute the second phase of the \textit{transformer} security status
by combining the results of the phase 1 (column ``T1 result'' in Table
\ref{table:rules_transformer1}) with column ``B.env result'' in Table
\ref{table:rules_build_environment}), shown in Table
\ref{table:rules_transformer2}. The column ``Transformer result'' shows the
resulting status for the \textit{transformer}.

\begin{table}
	\caption{Rules for transformer phase 2}
	\label{table:rules_transformer2}
	\centering
	\begin{tabular}[pos]{ |c|c|c|}

		\hline
		\textbf{T1 result} & \textbf{B.env result} & \textbf{Transformer result} \\
		\hline
		safe               & safe                  & safe                        \\
		\hline
		vulnerable         & safe                  & vulnerable                  \\
		\hline
		malicious          & safe                  & malicious                   \\
		\hline
		safe               & compromised           & malicious                   \\
		\hline
		vulnerable         & compromised           & malicious                   \\
		\hline
		malicious          & compromised           & malicious                   \\
		\hline
	\end{tabular}
\end{table}

We now introduce the log model which captures the threat propagation
contributions for the software supply chain.

\section{The Log Model}
\label{sec:logmodel}
We propose the \textit{log model} to capture information about the build
process of a software product. We record the actions that took place, the
artifacts that were involved, and the choices that were made within a software
supply chain for the building of a specific software product. We also capture
the topology of the data flows that took place.

We define the \textit{log model} $G_{LM}$, by adding vertex types, edge types
and properties to a standard property graph model $G$ \cite{Bonifati2018}. We
recall it here for referencing purposes:

\begin{definition}[Log Model]
	Let $L$ be a finite set of \textit{labels}, $K$ be a set of \textit
	{property keys} and $N$ be a set of \textit{values}.
	Let $G_{LM} = \left( V, E, \eta, \lambda, \nu \right)$ be the log
	model for a software supply chain, where:

	\begin{itemize}
		\item  $V$ is a finite set of vertices;
		\item  $E$ is a finite set of edges such that
		      $E \cap V = \emptyset$;
		\item  $\eta \colon E \mapsto V \times V$ is a total function that
		      maps each edge to an ordered pair of vertices;
		\item  $\lambda \colon V \cup E \mapsto P \left( L \right)$ is a
		      function that assigns to each \textit{vertex} and
		      \textit{edge} a finite set of \textit{labels};
		\item  $\nu \colon \left( V \cup E \right) \times K \mapsto N$ is a
		      partial function that assigns property values to elements,
		      such that the set of domain values where $\nu$ is defined is finite.
	\end{itemize}

\end{definition}

We constrain a general property graph to the set of vertices type in which we
are interested in.

\begin{definition}[Vertex types]
	Let $L \left( V \right)$ be the set of labels for $V$, such that:
	$L \left( V \right) \subseteq $ \{ \textit{softwareArtifact},
	\textit{transformer}, \textit{host}, \textit{buildEnvironment} \}

\end{definition}

The vertex types defined by those labels are:

\begin{enumerate}
	\item \textbf{softwareArtifact} - see definition \ref{def:software_artifact};

	\item \textbf{transformer} - see definition \ref{def:transformer};

	\item \textbf{host} - see definition \ref{def:host};

	\item \textbf{buildEnvironment} - see definition \ref{def:build_environment}.
\end{enumerate}

Let $S \in V$ be the set of all \textit{softwareArtifact} vertices, $T \in V$
be the set of all \textit{transformer} vertices, $H \in V$ be the set of all
\textit{host} vertices, and $B \in V$ be the set of all
\textit{buildEnvironment} vertices.

We further constrain the property graph $G_{LM}$ to contain only the following
edge types:

\begin{definition}[Edge types]
	Let $L \left( E \right)$ be the set of labels for $E$, such that:
	$	L \left( E \right) \subseteq$ \{
	\textit{hosted, executed, wasInputTo, wasPresent, generated,
		wasPublishedTo, transferred} \}
\end{definition}

The edges are all directed and their labels represent their types as explained
bellow:

\begin{enumerate}
	\item \textbf{hosted} - an edge that designates which host provided a
	      build environment. It connects a \textit{host} vertex to an
	      \textit{buildEnvironment} vertex;
	\item \textbf{executed} - an edge that specifies which build
	      environment was used to generate the build. It connects an
	      \textit{buildEnvironment} vertex to a \textit{transformer} vertex.
	\item \textbf{wasInputTo} - an edge that specifies the input of a
	      transformation. It connects a \textit{softwareArtifact} vertex to a
	      \textit{transformer} vertex.
	\item \textbf{wasPresent} - an edge that specifies what software
	      artifacts were present in the	\textit{buildEnvironment} and at
	      the host when the build process was executed. It connects a
	      \textit{softwareArtifact} vertex to an \textit{buildEnvironment} or to
	      a \textit{host} vertex.
	\item \textbf{generated} - an edge that specifies the results of a
	      transformation. It connects a \textit{transformer} vertex to a
	      \textit{softwareArtifact} vertex.
	\item \textbf{wasPublishedTo} - an edge that designates where generated
	      artifacts were published to be consumed by others. It connects a
	      \textit{softwareArtifact} to a \textit{host} vertex.
	\item \textbf{transferred} - an edge that designates where the artifacts
	      came from. It connects a \textit{host} vertex to a
	      \textit{softwareArtifact} vertex.

\end{enumerate}

The edges connection rules are summarized in Table \ref{table:edge_types}.

\begin{table}
	\caption{Edge types and the combination of vertices allowed}
	\label{table:edge_types}
	\centering
	\begin{tabular}[pos]{ |c|c|c|}

		\hline
		\textbf{Edge type} & \textbf{Source vertices} & \textbf{Target vertices} \\
		\hline
		hosted             & host                     & buildEnvironment         \\
		\hline
		executed           & buildEnvironment         & transformer              \\
		\hline
		wasInputTo         & softwareArtifact         & transformer              \\
		\hline
		wasPresent         & softwareArtifact         & buildEnvironment, host   \\
		\hline
		generated          & transformer              & softwareArtifact         \\
		\hline
		wasPublishedTo     & softwareArtifact         & host                     \\
		\hline
		transferred        & host                     & softwareArtifact         \\
		\hline

	\end{tabular}
\end{table}

\begin{figure}[h]
	\centering
	\includegraphics[width=0.5\textwidth]{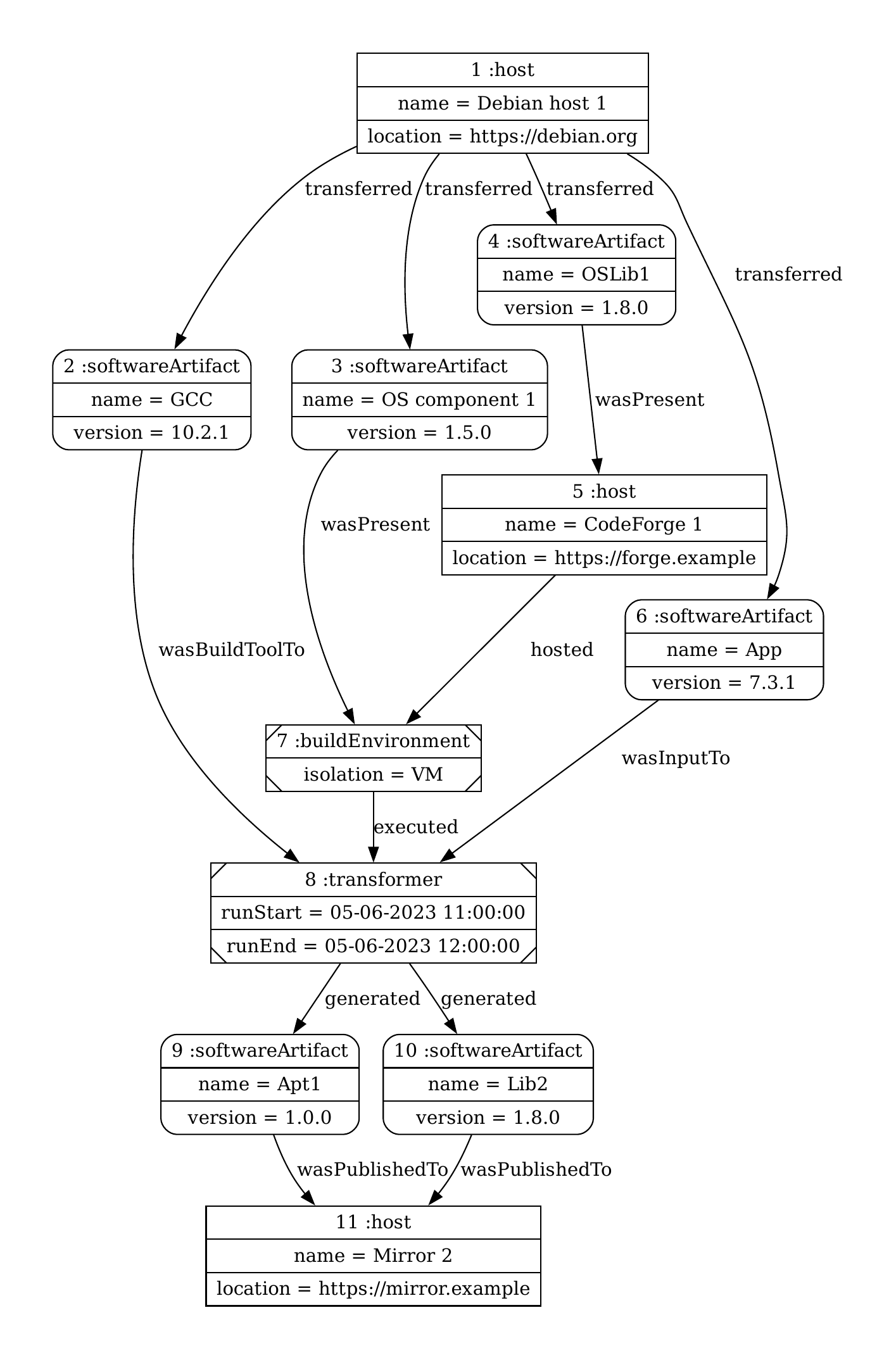}
	\caption{An example of the Log Model}
	\label{fig:1-log-model}
\end{figure}

An example of a \textit{log model} property graph $G_{LM}$ is rendered in
figure \ref{fig:1-log-model}. \textit{Hosts} (vertices 1, 5 and 11) are
rendered with rectangle shapes, \textit{software artifacts} (vertices 2, 3, 4,
6, 9 and 10) are rendered in rounded corners. \textit{Build environments}
(vertex 7) and \textit{transformers} (vertex 8) are rendered with diagonal
corners. For brevity we show only some of the properties for each element of
the graph. Element \#4, a \textit{software artifact}, \textit{was present} in
element \#5, a \textit{host}, which \textit{hosted} element \#7, a
\textit{build environment}, which used a virtual machine (VM) to execute
element \#8, a \textit{transformer}. Element \#2, a \textit{software artifact},
\textit{was a build tool} to the \textit{transformer}, while element \#6, a
\textit{software artifact}, \textit{was input} to the same
\textit{transformer}, which \textit{generated} elements \#9 and \#10,
\textit{software artifacts}. Finally, those artifacts \textit{were published}
to element \#11, a \textit{host}. In this example, those actions and
characteristics were observed during the building run of element \#8, the
\textit{transformer}.

\section{Security Status Calculus}
\label{sec:analysis}

This sections defines how to calculate the software security status of the
elements of a software supply chain, as captured by the \textit{log model}. For
any starting element, it is necessary to consider all the other elements among
all the paths that lead to it.

We introduce the following notation for typing purposes: $S$, $H$, $T$, and $B$
stand for \textit{softwareArtifact}, \textit{host}, \textit{transformer}, and
\textit{buildEnvironment} vertices, respectively. We introduce the set of
status $Q=\{safe, vulnerable, malicious\}$, to represent the status of both
hosts and software artifacts (see \ref{def:sw_security_state}). If a $host$ has
the label $malicious$ it means it is equivalent to being compromised. For a
given set $X$, $\cal{P}(X)$ designates the subsets of $X$. The list of security
status $Q$ is ordered and defined for any $X\in\cal{P}(Q)$ and $max(X)$ returns
the singleton equal to the greatest label in $X$ or the empty set if $X$ is
empty. The function \textit{parentsOf($v \colon V$)} returns a set of parent
vertices of $V$, possibly empty. Let \textit{nil} be a symbol indicating no
value. The sets $\chi_{SV}$ and $\chi_{SM}$, and the \textit{log model} are
considered to be globally available to be consulted from all functions.

\label{def:results}
\begin{definition}[Results] Let a calculated state of the security calculus be
	the	5-ple $Res = (q, S_V, S_M, H_V, H_M )$ , where:

	\begin{itemize}
		\item $q \in Q$ is the security status of the desired element;
		\item $S_V$ is a set of \textit{vulnerable}
		      \textit{softwareArtifact} vertices;
		\item $S_M$ is a set of \textit{malicious}
		      \textit{softwareArtifact} vertices
		\item $H_V$ is a set of \textit{vulnerable}
		      \textit{host} vertices
		\item $H_M$ is a set of \textit{malicious}
		      \textit{host} vertices

	\end{itemize}
\end{definition}

All functions that can be called for getting an element status in the algorithm
listings return a $Res$ 5-ple so all the \textit{host} and
\textit{softwareArtifact} vertices security states from the origins to the
selected element in the \textit{log model} will be available (see
\ref{sec:usecases} for use cases).

We will use the following notation for operating with instances of type $Res$.
For any element $r = (x_1,x_2,\dots, x_5)$:

\begin{itemize}
	\item $r[i]$ means read the $x_i$ value;
	\item $r[i] := y$ means attribute value $y$ to element $x_i$;
\end{itemize}

We will use the special helper function \textit{merge} to define the first
element $q$ and to apply the individual unions of all other elements of the
5-tuples of type $Res$.

\begin{definition}[merge] Let $merge \colon Q \times Res \mapsto Res$ be
	a function that merges two instances of 5-ples $Res$, such that:
	$merge(q, X, Y) = (q, x_2 \cup y_2, x_3 \cup y_3, \dots,  x_5 \cup y_5)$

\end{definition}

The entry point for the algorithm for calculating the status of a
\textit{softwareArtifact} is shown in Listing~\ref{list:getSoftwareStatus}.
Later we will present the functions that gather information about other vertex
types. We combine the initial condition of vertex $s$ ($initialSoftwareStatus$
in Listing~\ref{list:initialSoftwareStatus}) with the recursively calculated
status of all parents of $s$.

\newtcbtheorem{algorithm2}{Listing}{pseudo/ruled, float}{list}

\pseudoset{
	ct-left=\texttt{/\!/}\,,ct-right=,ctfont=\color{black!65}
}

\begin{algorithm2}[]{}{getSoftwareStatus}

\begin{pseudo}[label=\small\arabic*, indent-mark]
	function \bf{getSoftwareStatus($s \colon S$, $r_0 \colon Res$)}: $Res$	\\+
		let $r \colon Res$ := copyOf($r_0$) \\	
		let $r_p \colon Res$ := getAllStatuses($s$, parentsOf($s$), $r_0$) \\
		let $q \colon Q$ := max(initialSoftwareStatus($s$), $r_p[1]$) \\
		
		\ct{add this sw to the corrensponding set} \\
		if $q$ =  \textit{vulnerable} then \\+
			$r[2]$ := $r[2] \cup s$ \\-
		if $q$ =  \textit{malicious} then \\+
			$r[3]$ := $r[3] \cup s$ \\-
		return merge($q$, $r$, $r_p$)\\-
\end{pseudo}

\end{algorithm2}

\begin{algorithm2}[]{}{initialSoftwareStatus}

	\begin{pseudo}[label=\small\arabic*, indent-mark]
		function \bf{initialSoftwareStatus}($s \colon S$): $Q$ \\+
			let $a$,$b$ := \id{safe},\id{safe}	 \\
			if $s \in \chi_{SV}$ then \\+
				$a$ := \id{vulnerable} \\-
			if $s \in \chi_{SM}$ then \\+
				$b$ := \id{malicious} \\-
			return max($a$, $b$) \\--
	\end{pseudo}

\end{algorithm2}

Given a set of vertices $V$ returned by the function $parentsOf$, we calculate
the final security status by combining all $V$ statuses, according to the
specific rules for each type of vertex. We do that by calling the specific
function as required, as shown in Listing~\ref{list:getAllStatuses}. When any
of those specific functions are called, they call the $parentsOf$ function
again, recursively visiting the remaining incoming edges of the graph. There is
an exceptional rule that will be presented later, which requires the knowledge
of the current analyzed software artifact $s \in S$ to calculate the status of
a \textit{host}.

\begin{algorithm2}[]{}{getAllStatuses}

	\begin{pseudo}[label=\small\arabic*, indent-mark]
		function \bf{getAllStatuses}($s : S$, $V_x \colon \cal{P}(V)$, $r_0 \colon Res$): $Res$	\\+
			let $r \colon Res$ := copyOf($r_0$) \\
			for each $v \in V_x$ \\+
				let $r_p$:$Res$ := getOneStatus(s, v, $r_0$) \\
				$r$ := merge(max($r[1], r_p[1]$), $r$, $r_p$) \\-
			return $r$ \\-
	\end{pseudo}

	\begin{pseudo}[label=\small\arabic*, indent-mark]
		function \bf{getOneStatus}($s \colon S$, $v \colon V$, $r_0 \colon Res$): $Res$		\\+
			
			if $v \neq \textit{nil}$ then \\+
				case typeOf($v$) of  \\+
					$S$: return getSoftwareStatus($v$, $r_0$)) \\
					$T$: return getTransformerStatus($v$, $r_0$)) \\
					$B$: return getBuildEnvironmentStatus($v$, $r_0$) \\
					$H$: return getHostStatus($s$, $v$, $r_0$) \\--
			\ct{$v$ doesn't exist or element not known} \\
			let $r$ := copyOf($r_0$) \\
			$r[1]$ := \id{malicious} \\
			return $r$
	\end{pseudo}

\end{algorithm2}

For a \textit{transformer} node we look at the incoming edges and separate the
source vertices into two sets: $B_{tools} \subseteq S$, the set of software
artifacts that are build tools, and $A \subseteq V$, the set of any other
vertices. For a \textit{transformer} to propagate a vulnerability downstream it
must come from a vulnerable component that it is used as a building block
(i.e., from a \textit{wasInputTo} edge). A vulnerable build tool, meanwhile, is
not copied to the generated components, so its security status is not
propagated. However, a \textit{malicious} software artifact that is used as a
build tool (e.g., a compiler, an archiver) can inject malicious code into the
generated artifacts~\cite{Thompson1984} and thus, it propagates the
\textit{malicious} security status. For this reason, we need to check if there
is \textit{malicious} status related to a building tool. If there is not, the
resulting security status will result exclusively from the calculations of the
other vertices. Listing~\ref{list:getTransformerStatus} shows the algorithm.

\begin{algorithm2}[]{}{getTransformerStatus}

	\begin{pseudo}[label=\small\arabic*, indent-mark]
		function \bf{getTransformerStatus($t \colon T$, $r_0 \colon Res$)}: $Res$ \\+
			let $B_{tools} \colon S$ := $\emptyset$ \\
			let $A \colon V$ := $\emptyset$ \\
			\ct{Separate building tools from other elements}\\
			for each $v \in $ parentsOf($t$) \\+
				if typeOf($v$) = $S$ $\wedge$ \\+
					 outgoing edge of $v$ = \id{wasBuildToolTo} then  \\+
					$B_{tools}$ := $B_{tools} \cup \{v\}$ \\-
				else \\+
					$A$ := $A \cup \{v\}$ \\--

			let $r_t \colon Res$ := getAllStatuses(\id{nil}, $B_{tools}$, $r_0$) \\
			let $r_a \colon Res$ := getAllStatuses(\id{nil}, $A$, $r_0$) \\

			\ct{consider build tools status only if malicious} \\
			let $q \colon Q$ := \id{safe} \\
			if \id{mailicous} = $r_t[1]$ then \\+
				$q$ := \id{mailicous} \\-

			\ct{Merge all elements} \\
			return merge(max($q$, $r_a[1]$), $r_t$, $r_a$) \\--
	\end{pseudo}

\end{algorithm2}

There are only two possible edges that end in a \textit{buildEnvironment}
vertex $b \in B$: \textit{hosted} and \textit{wasPresent}. Since nothing is
copied from $b$ by a \textit{transformer}, the \textit{vulnerable} security
status is never propagated from $b$. However, both a \textit{compromised} host
or a \textit{malicious} software artifact that was present inside $b$ could
inject \textit{malicious} code into the other software artifacts being
processed by the \textit{transformer} that executed inside $b$.
Listing~\ref{list:getBuildEnvironmentStatus} shows the algorithm that
calculates de security status for $b$.

\begin{algorithm2}[]{}{getBuildEnvironmentStatus}

	\begin{pseudo}[label=\small\arabic*, indent-mark]
		function \bf{getBuildEnvironmentStatus}($b \colon B$, $r_0 \colon Res$): $Res$ \\+
			let $r \colon Res$ := getAllStatuses(\id{nil}, parentsOf($b$), $r_0$) \\
			if $r[1]$ = \id{malicious} then \\+
				return $r$ \\-
			\ct{Only malicious threats propagate} \\	
			$r[1]$ = \id{safe} \\
			return $r$\\-
	\end{pseudo}

\end{algorithm2}

The security calculus of a \textit{host} is the combination of its initial
security status (i.e., whether or not it was known as \textit{compromised}
during the execution of the \textit{buildEnvironment}) with all the possible
security risks that its parent vertices might bring. The propagation rules
depend on the role the \textit{host} is performing.

When the \textit{host} is executing a \textit{buildEnvironment} (i.e., it is
the parent of a \textit{buildEnvironment}) vertex, the only possibility of
propagating threats is when it is compromised and thus, it is able to inject
malicious code. For that to happen the \textit{host} has to be attacked or it
has to execute some \textit{softwareArtifact} that is \textit{malicious} and
was connected by a \textit{wasPresent} edge. A \textit{vulnerable} software
that is among the supporting software from a host, but is not a building block
type of dependency (i.e., it is not copied or transformed by the
\textit{transformer}) has no effect on the propagation of security status from
the \textit{host} on this role.

When the \textit{host} is performing the role of a storage from which to
transfer one or more software artifacts, the propagation rules also must
account for the children (e.g., who was using each software artifact). Since in
this role software artifacts are copied from the \textit{host} to be processed
by the \textit{transformer}, the security status propagation must account for
this usage. So the algorithm consider the special case where a software
artifact was transferred and it is indicated in the algorithm by specifying the
child $s$. Listing~\ref{list:getHostStatus} shows the algorithm that calculates
the security status for a \textit{host} $h$, considering the possibility where
the child vertex may be the \textit{softwareArtifact} $s$.

\begin{algorithm2}[]{}{getHostStatus}

	\begin{pseudo}[label=\small\arabic*, indent-mark]
		function \bf{getHostStatus($s \colon S$, $h \colon H$, $r_0 \colon Res$)}: $Res$ \\+
			let $r \colon Res$ = copyOf($r_0$) \\
			if $h \in \chi_H$ then \\+
				$r[1]$ := \textit{malicious} \\-
			let $P_h \colon \cal{P}(V)$ :=  parentsOf($h$) \\
			let $P_{os} \colon \cal{P}(V)$ :=  $\{v| v.edge = \textit{wasPresent} \wedge v \in P_h\}$ \\
			let $r_{os}$ := getAllStatuses($s$, $P_{os}$, $r_0$) \\
			$r$ := merge(max(r[1], $r_{os}$[1]), $r$, $r_{os}$) \\
			if $r[1]$ = \id{vulnerable} then \\+
				$r[4]$ := $r[4] \cup h$ \ct{saves h in the set} \\
				$r[1]$ := \textit{safe} \ct{can't propagate vuln. here}\\-
			let $r_s \colon Res$ := copyOf($r_0$) \\
			if $s \neq \textit{nil} \wedge P_h \neq \emptyset$ then \\+
				\ct{Look for a copy of $s$ that was}  \\
				\ct{transferred to the host} \\
				let $p$ := $x | x \in P_h, id(x) = id(s)$ \\
				\ct{Sw building block contribution} \\
				if $p \neq \textit{nil}$ then \\+
					$r_s$ = getOneStatus($s$, $p$, $r_0$) \\
					\ct{if h not compromised, $r_s$ (re)defines status} \\
					let $q : Q$ := \textit{malicious} \\
					if $r[1] \neq \textit{malicious}$ then \\+
						$q$ := $r_s[1]$ \\-
					$r$ = merge($q$, $r$, $r_s$) \\--

			if $r[1]$ = \textit{malicious} \\+
				\ct{saves h in the compromised set} \\
				$r[5]$ := $r[5] \cup h$ \\-
			return $r$ \\-
	\end{pseudo}

\end{algorithm2}

\section{Use Cases}
\label{sec:usecases}

We present use cases where the \textit{log model} can be used to give first
insights on the \SSC security status and additional insights when more security
information becomes available.

\begin{figure}[h]
	\centering
	\includegraphics[width=0.5\textwidth]{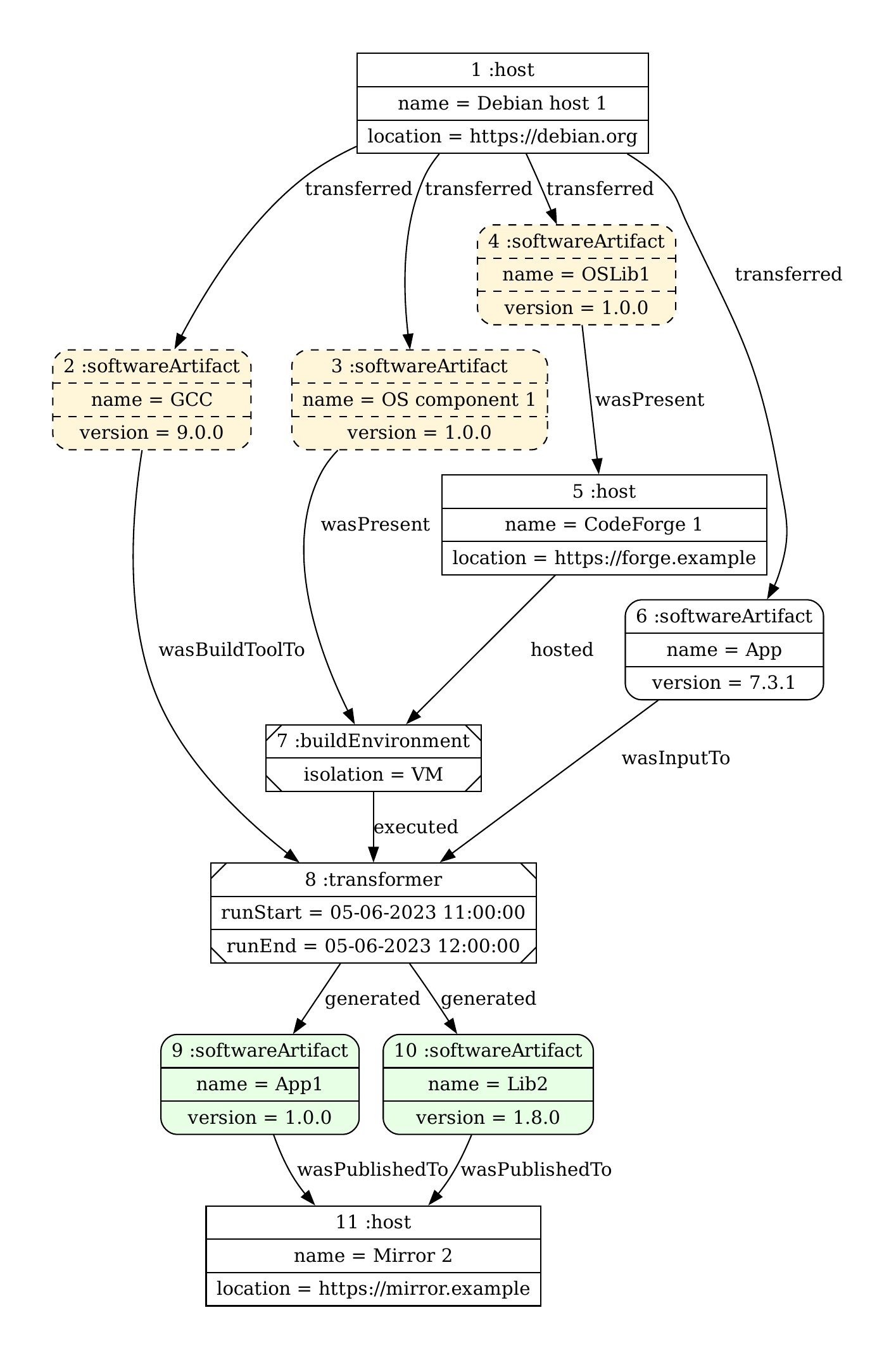}
	\caption{Vulnerability in some components (shown by dashed lines),
		don't propagate to software products.}
	\label{fig:2-log-model}
\end{figure}

\begin{usecase}[Vulnerable components, but safe software products]
	Figure \ref{fig:2-log-model} shows the \SSC for software artifacts App1
	(i.e., vetertex 9) and Lib2 (i.e., vertex 10). Let's suppose that just after
	our building and publishing of those artifacts, it was disclosed by
	the software community that the artifacts GCC v9.0.0 (i.e., vertex 2), OS
	component 1 	v1.0.0 (i.e., vertex 3), and OSLib1 v1.0.0 (i.e., vertex 4) are
	all vulnerable to attacks. Thus $ \{ v_3, v_4, v_5 \} \in \chi_V$.
	By applying function
	\textbf{getSoftwareStatus} (Listing~\ref{list:getSoftwareStatus})
	on vertices 9 and 10, we are 	reassured that even though there were
	some vulnerable artifacts	in the \SSC, vertices 9 and 10 still have the
	\textit{safe} security
	status. Additionally, we might want to verify if there are known attacks
	in the wild that exploit those vulnerabilities of vertices 2, 3 and 4. If
	it is so, we might want to request an audit of the host CodeForge 1 (i.e.,
	vertex 5), because its risk of future compromise grows when exploits are
	already observed in the wild.
\end{usecase}

\begin{figure}[h]
	\centering
	\includegraphics[width=0.5\textwidth]{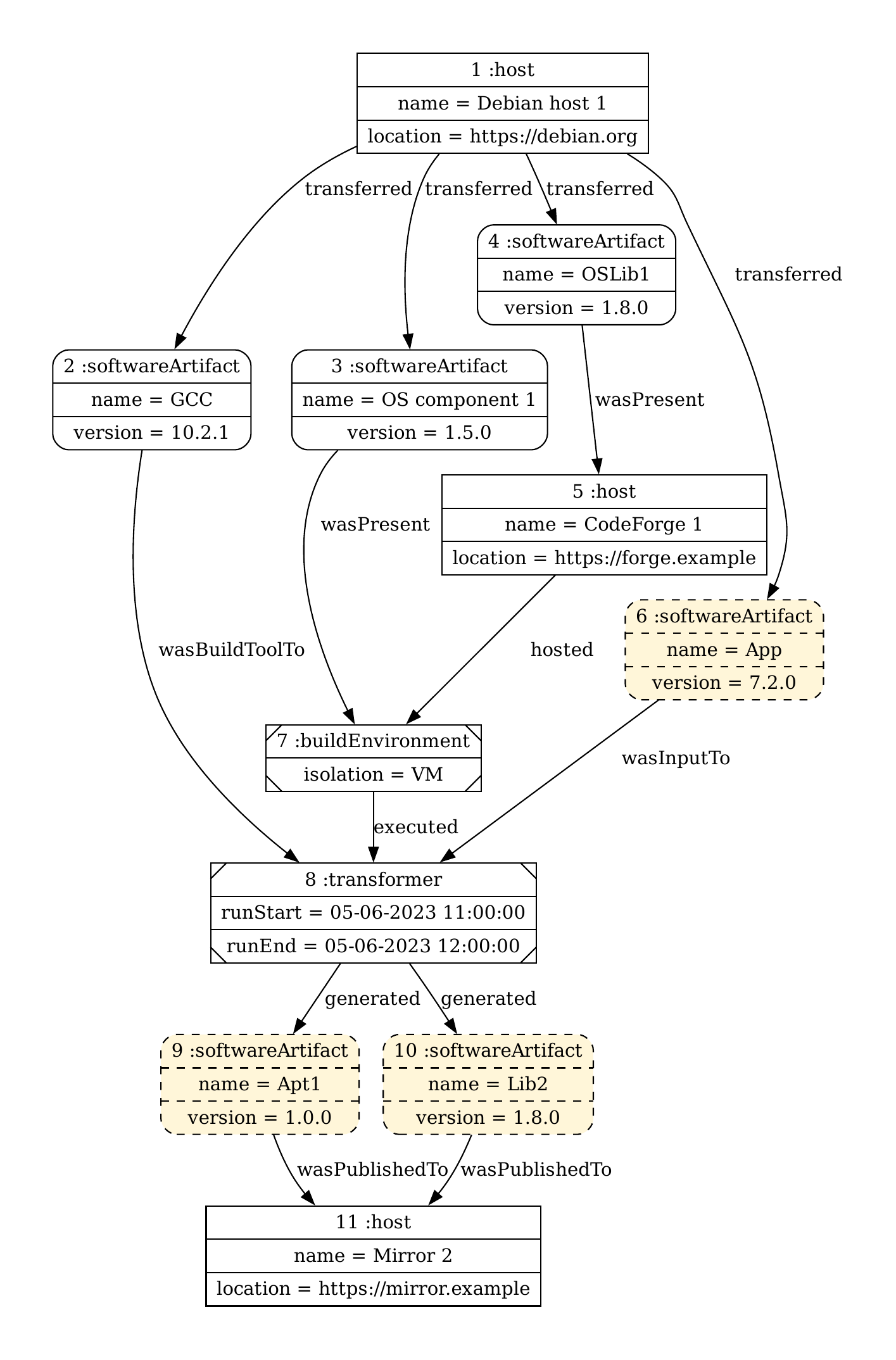}
	\caption{A vulnerability in component \#9 propagates to
		the software products.	(Vulnerabilities shown by dashed lines)}
	\label{fig:3-log-model}
\end{figure}

\begin{usecase}[Vulnerable component causes vulnerable software products]
	Figure \ref{fig:3-log-model} shows the \SSC  for software artifacts App1
	(i.e., vertex 9) and Lib2 (i.e., vertex 10). In this case, there was
	the late disclosure that the artifact App v7.2.0 (i.e., vertex 6) is
	vulnerable. Therefore, $ \{ v_6 \} \in \chi_V$. By applying function
	\textbf{getSoftwareStatus} (Listing~\ref{list:getSoftwareStatus})
	on vertices 9 and 10, we see that both also have the
	\textit{vulnerable} security status. Additionally, we can analyze how
	to patch or upgrade vertex 6 in such a way as to remove its
	vulnerability and prepare a new release of the software products
	App1 and Lib2.
\end{usecase}

\begin{figure}[h]
	\centering
	\includegraphics[width=0.5\textwidth]{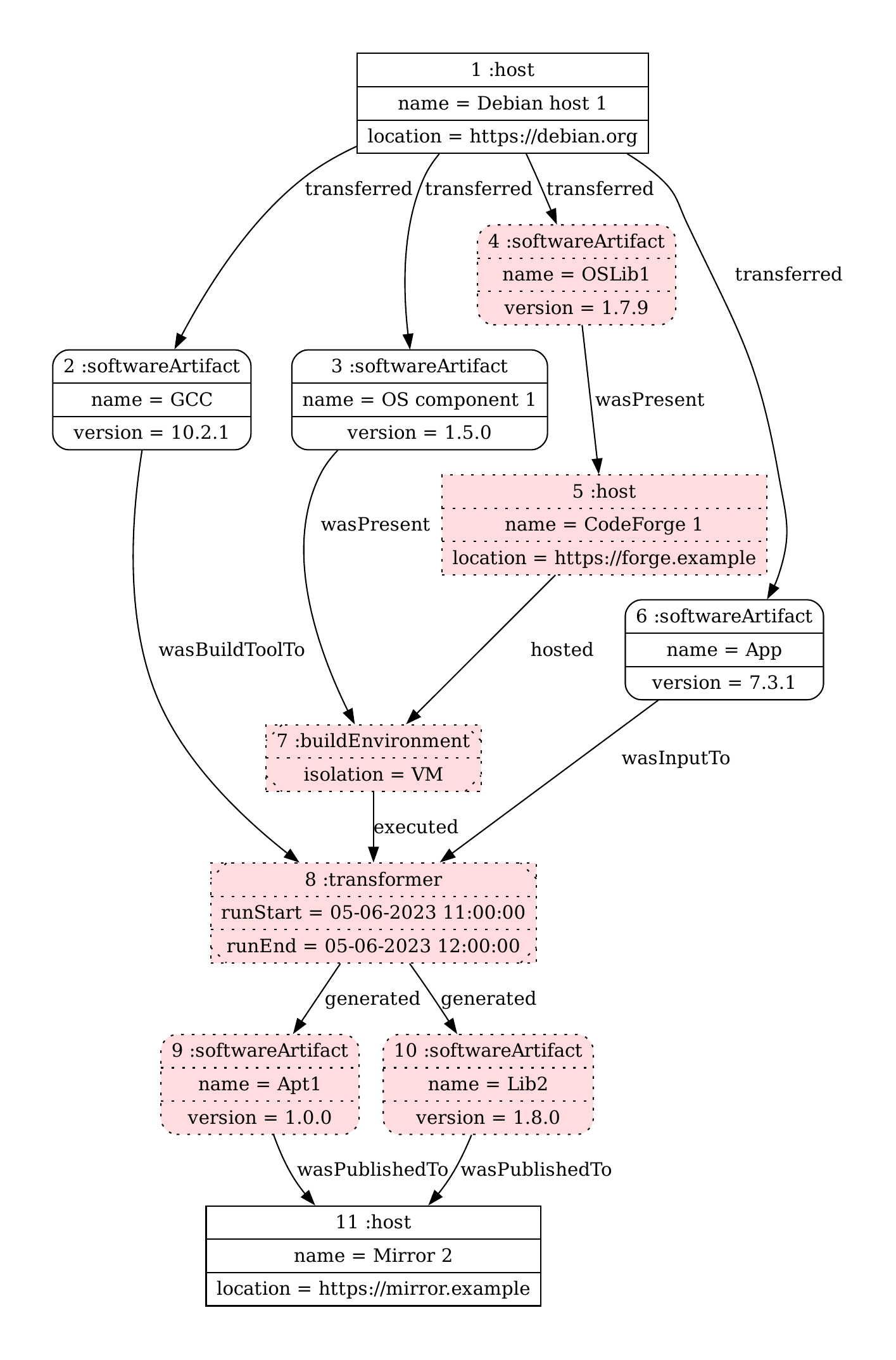}
	\caption{The propagation of malicious status of component
		\#4  (shown by dotted lines)}
	\label{fig:4-log-model}
\end{figure}

\begin{usecase}[Malicious software component compromises \SSC]
	Figure \ref{fig:4-log-model} shows the \SSC  for software artifacts
	App1 (i.e., vertex 9) and Lib2 (i.e., vertex 10). After the build
	is finished, it comes to attention that there was an attack which
	resulted of malicious code being present in the component OSLib1 v1.7.9
	(i.e., vertex 4). So $\{ v_4 \} \in \chi_M$. By applying function
	\textbf{getSoftwareStatus}  (Listing~\ref{list:getSoftwareStatus})
	on vertices 9 and 10 we are	able to discover that both can be considered malicious (they	have a high risk of being malicious). Furthermore, the result of the 	calculation (see \ref{def:results}) shows that the set of calculated compromised hosts $H_M = \{5\}$, and the set of malicious software artifacts $S_M = \{4, 9, 10\}$. We could request a security audit for the host depicted in vertex 5 and request the removal of the software products App1 and Lib2 from the host depicted in	vertex 11, to prevent them to contaminate other parties.
\end{usecase}

\section{Conclusion}
\label{sec:conclusion}

The security of software supply chains is a growing concern and the task of
assessing the threat risks associated with any given software product is non
trivial.

The main contribution presented in this paper is the \textit{log model}, an
approach to keep the history of software supply chain activities that occurred
during the production of a software artifact. The model allows the use of rules
for tracing the threat propagation among the software supply chain elements.
The application of all the rules combined with external knowledge of host
compromises and vulnerabilities disclosures can help practitioners assess the
security risks for software artifacts, hosts, and other elements of a \SSC.
Moreover, the information obtained can be used to help determine the best
course of action for corrective measures, if required (e.g., what software
components to patch or upgrade, what hosts to further look for security
breaches). Finally, by mapping all elements in the software supply chain that
may be vulnerable, the model allows practitioners to watch more closely for
related exploits found in the wild and take preventive action.

For future research, it could be helpful to augment the threat propagation
rules by considering temporal markers for disclosed compromises and
vulnerabilities. The current proposal assumes that the set of software
vulnerabilities and compromises and the set of host compromises occurred in an
interval of time that overlapped with the time interval where software building
took place in the software supply chain. By including fine grained temporal
handling in the calculation, the model could help avoid false positives (e.g.,
even if there was a compromised host involved in the process, its compromise
could have happened after its participation was already finished in a build,
leaving other transformations going on other unrelated parts of the software
supply chain).

Another area of research is the possibility of adding other external measures
to the propagation calculations, such as reputation of a resource, and
quantifying the threats levels in each category by range of values.


\begin{thebibliography}{10}

\bibitem{Ohm2020}
M.~Ohm, H.~Plate, A.~Sykosch, and M.~Meier, ``Backstabber's knife collection: A
  review of open source software supply chain attacks,'' in {\em Detection of
  Intrusions and Malware, and Vulnerability Assessment}, pp.~23--43, Springer
  International Publishing, 2020.

\bibitem{Ladisa2022}
P.~Ladisa, H.~Plate, M.~Martinez, and O.~Barais, ``Taxonomy of attacks on
  open-source software supply chains,'' 2022.

\bibitem{Peisert2021}
S.~Peisert, B.~Schneier, H.~Okhravi, F.~Massacci, T.~Benzel, C.~Landwehr,
  M.~Mannan, J.~Mirkovic, A.~Prakash, and J.~B. Michael, ``Perspectives on the
  {SolarWinds} incident,'' {\em {IEEE} Security \& Privacy}, vol.~19,
  pp.~7--13, mar 2021.

\bibitem{ghadge2020managing}
A.~Ghadge, M.~Wei{\ss}, N.~D. Caldwell, and R.~Wilding, ``Managing cyber risk
  in supply chains: A review and research agenda,'' {\em Supply Chain
  Management: An International Journal}, vol.~25, no.~2, pp.~223--240, 2020.

\bibitem{hammi2023security}
B.~Hammi, S.~Zeadally, and J.~Nebhen, ``Security threats, countermeasures, and
  challenges of digital supply chains,'' {\em ACM Comput. Surv.}, vol.~55, jul
  2023.

\bibitem{Harutyunyan2020}
N.~Harutyunyan, ``Managing your open source supply chain-why and how?,'' {\em
  Computer}, vol.~53, pp.~77--81, jun 2020.

\bibitem{Vu2021}
D.~L. Vu, I.~Pashchenko, F.~Massacci, H.~Plate, and A.~Sabetta, ``Lastpymile
  replication package,'' 2021.

\bibitem{Abate2020DepSolving}
P.~Abate, R.~D. Cosmo, G.~Gousios, and S.~Zacchiroli, ``Dependency solving is
  still hard, but we are getting better at it,'' in {\em 27th {IEEE}
  International Conference on Software Analysis, Evolution and Reengineering,
  {SANER} 2020, London, ON, Canada, February 18-21, 2020} (K.~Kontogiannis,
  F.~Khomh, A.~Chatzigeorgiou, M.~Fokaefs, and M.~Zhou, eds.), pp.~547--551,
  {IEEE}, 2020.

\bibitem{Decan2019DepNetworks}
A.~Decan, T.~Mens, and P.~Grosjean, ``An empirical comparison of dependency
  network evolution in seven software packaging ecosystems,'' {\em Empir.
  Softw. Eng.}, vol.~24, no.~1, pp.~381--416, 2019.

\bibitem{German2017LicenseInconsistencies}
Y.~Wu, Y.~Manabe, T.~Kanda, D.~M. Germ{\'{a}}n, and K.~Inoue, ``Analysis of
  license inconsistency in large collections of open source projects,'' {\em
  Empir. Softw. Eng.}, vol.~22, no.~3, pp.~1194--1222, 2017.

\bibitem{Ombredanne2020}
P.~Ombredanne, ``Free and open source software license compliance: Tools for
  software composition analysis,'' {\em Computer}, vol.~53, pp.~105--109, oct
  2020.

\bibitem{Imtiaz2021}
N.~Imtiaz, S.~Thorn, and L.~Williams, ``A comparative study of vulnerability
  reporting by software composition analysis tools,'' in {\em Proceedings of
  the 15th {ACM} / {IEEE} International Symposium on Empirical Software
  Engineering and Measurement ({ESEM})}, {ACM}, oct 2021.

\bibitem{Schneier1999AttackTrees}
B.~Schneier, ``Attack trees,'' {\em Dr. Dobb’s journal}, vol.~24, no.~12,
  pp.~21--29, 1999.

\bibitem{8240774}
V.~Mavroeidis and S.~Bromander, ``Cyber threat intelligence model: An
  evaluation of taxonomies, sharing standards, and ontologies within cyber
  threat intelligence,'' in {\em 2017 European Intelligence and Security
  Informatics Conference (EISIC)}, pp.~91--98, 2017.

\bibitem{7180277}
B.~A. Sabbagh and S.~Kowalski, ``A socio-technical framework for threat
  modeling a software supply chain,'' {\em IEEE Security \& Privacy}, vol.~13,
  no.~4, pp.~30--39, 2015.

\bibitem{duman2019modeling}
O.~Duman, M.~Ghafouri, M.~Kassouf, R.~Atallah, L.~Wang, and M.~Debbabi,
  ``Modeling supply chain attacks in iec 61850 substations,'' in {\em 2019 IEEE
  International Conference on Communications, Control, and Computing
  Technologies for Smart Grids (SmartGridComm)}, pp.~1--6, IEEE, 2019.

\bibitem{yamaguchi2014modeling}
F.~Yamaguchi, N.~Golde, D.~Arp, and K.~Rieck, ``Modeling and discovering
  vulnerabilities with code property graphs,'' in {\em 2014 IEEE Symposium on
  Security and Privacy}, pp.~590--604, IEEE, 2014.

\bibitem{10123571}
A.~M. Mir, M.~Keshani, and S.~Proksch, ``On the effect of transitivity and
  granularity on vulnerability propagation in the maven ecosystem,'' in {\em
  2023 IEEE International Conference on Software Analysis, Evolution and
  Reengineering (SANER)}, pp.~201--211, 2023.

\bibitem{perl2015vccfinder}
H.~Perl, S.~Dechand, M.~Smith, D.~Arp, F.~Yamaguchi, K.~Rieck, S.~Fahl, and
  Y.~Acar, ``{VCCFinder}: Finding potential vulnerabilities in open-source
  projects to assist code audits,'' in {\em Proceedings of the 22nd {ACM}
  {SIGSAC} Conference on Computer and Communications Security, Denver, CO, USA,
  October 12-16, 2015} (I.~Ray, N.~Li, and C.~Kruegel, eds.), pp.~426--437,
  {ACM}, 2015.

\bibitem{8009930}
P.~Behnamghader, R.~Alfayez, K.~Srisopha, and B.~Boehm, ``Towards better
  understanding of software quality evolution through commit-impact analysis,''
  in {\em 2017 IEEE International Conference on Software Quality, Reliability
  and Security (QRS)}, pp.~251--262, 2017.

\bibitem{Zahan2023}
N.~Zahan, E.~Lin, M.~Tamanna, W.~Enck, and L.~Williams, ``Software bills of
  materials are required. are we there yet?,'' {\em {IEEE} Security \&
  Privacy}, vol.~21, pp.~82--88, mar 2023.

\bibitem{Xia2023}
B.~Xia, T.~Bi, Z.~Xing, Q.~Lu, and L.~Zhu, ``An empirical study on software
  bill of materials: Where we stand and the road ahead,'' Jan. 2023.

\bibitem{stewart2010software}
K.~Stewart, P.~Odence, and E.~Rockett, ``Software package data exchange (spdx)
  specification,'' {\em IFOSS L. Rev.}, vol.~2, p.~191, 2010.

\bibitem{7774522}
S.~S. Alqahtani, E.~E. Eghan, and J.~Rilling, ``Sv-af — a security
  vulnerability analysis framework,'' in {\em 2016 IEEE 27th International
  Symposium on Software Reliability Engineering (ISSRE)}, pp.~219--229, 2016.

\bibitem{Chang2011}
Y.-Y. Chang, P.~Zavarsky, R.~Ruhl, and D.~Lindskog, ``Trend analysis of the
  {CVE} for software vulnerability management,'' in {\em 2011 {IEEE} Third
  Int{\textquotesingle}l Conference on Privacy, Security, Risk and Trust and
  2011 {IEEE} Third Int{\textquotesingle}l Conference on Social Computing},
  {IEEE}, oct 2011.

\bibitem{Orman2003}
H.~Orman, ``The morris worm: a fifteen-year perspective,'' {\em {IEEE} Security
  \& Privacy}, vol.~1, pp.~35--43, sep 2003.

\bibitem{Maniriho2022}
P.~Maniriho, A.~N. Mahmood, and M.~J.~M. Chowdhury, ``A study on malicious
  software behaviour analysis and detection techniques: Taxonomy, current
  trends and challenges,'' {\em Future Generation Computer Systems}, vol.~130,
  pp.~1--18, may 2022.

\bibitem{Liao2013}
H.-J. Liao, C.-H.~R. Lin, Y.-C. Lin, and K.-Y. Tung, ``Intrusion detection
  system: A comprehensive review,'' {\em Journal of Network and Computer
  Applications}, vol.~36, pp.~16--24, jan 2013.

\bibitem{Bonifati2018}
A.~Bonifati, G.~Fletcher, H.~Voigt, and N.~Yakovets, {\em Querying Graphs}.
\newblock Springer International Publishing, 2018.

\bibitem{Thompson1984}
K.~Thompson, ``Reflections on trusting trust,'' {\em Communications of the
  {ACM}}, vol.~27, pp.~761--763, aug 1984.

\end{thebibliography}
\end{document}